\documentstyle[prd,aps,epsf,aps12,psfig,amssymb,amsmath,amsthm]{revtex}

\tighten
\begin{document}
\newcommand{\half}{\mbox{$\textstyle \frac{1}{2}$}}
\newcommand{\quat}{\mbox{$\textstyle \frac{1}{4}$}}
\newcommand{\octa}{\mbox{$\textstyle \frac{1}{8}$}}

\title{Relaxation of quantum states under energy perturbations}

\author{Dorje C. Brody$^{*}$, Lane P. Hughston$^{\dagger}$ and
Joanna Syroka$^{*}$}

\address{$*$Blackett Laboratory, Imperial College,
London SW7 2BZ, UK }

\address{$\dagger$Department of Mathematics, King's College London,
Strand, London WC2R 2LS, UK }

\date{\today}
\maketitle

\begin{abstract}
The energy-based stochastic extension of the Schr\"odinger
equation is perhaps the simplest mathematically rigourous and
physically plausible model for the reduction of the wave function.
In this article we apply a new simulation methodology for the
stochastic framework to analyse formulae for the dynamics of a
particle confined to a square-well potential. We consider the
situation when the width of the well is expanded instantaneously.
Through this example we are able to illustrate in detail how a
quantum system responds to an energy perturbation, and the
mechanism, according to the stochastic evolutionary law, by which
the system relaxes spontaneously into one of the stable
eigenstates of the Hamiltonian. We examine in particular how the
expectation value of the Hamiltonian and the probability
distribution for the position of the particle change in time. An
analytic expression for the typical timescale of relaxation is
derived. We also consider the small perturbation limit, and
discuss the relation between the stochastic framework and the
quantum adiabatic theorem.
\end{abstract}

\section{Introduction}

Consider a quantum system for which the Hamiltonian undergoes a
sudden change ${\hat h}\rightarrow{\hat H}$ at time $t=0$ in such
a manner that no mechanical `work' is done on or by the system.
Here ${\hat h}$ denotes the Hamiltonian before the change, and
${\hat H}$ denotes the Hamiltonian after the change.

If the initial wave function of the system is given by
$\psi_0(x)$, the probability that the wave function $\psi_t(x)$ at
a later time $t>0$ will be found in the $n$-th energy eigenstate
$\chi_n(x)$ of the new Hamiltonian ${\hat H}$ is determined by the
transition amplitude between $\psi_0(x)$ and $\chi_n(x)$. The
assumption that no work is done on or by the system during the
sudden change in the Hamiltonian implies that the expectation of
the Hamiltonian is preserved under such a transformation. That is
to say,  the relation
\begin{eqnarray}
\int \psi_0^*(x) {\hat h} \psi_0(x) {\rm d}x  = \int \psi_t^*(x)
{\hat H} \psi_t(x) {\rm d}x , \label{eq:1}
\end{eqnarray}
holds even though the Hamiltonian itself has changed.

Just after the application of the perturbation, the wave function
can be represented by an expansion of $\psi_0(x)$ in terms of the
eigenfunctions $\chi_n(x)$ of ${\hat H}$. The state then evolves
according to the unitary law governed by ${\hat H}$ in such a way
that the expectation value of ${\hat H}$ is preserved. According
to the unitary law, the wave function $\psi_t(x)$ will remain in a
state of superposition of the eigenmodes of ${\hat H}$, and this
superposition will linger indefinitely in time. Indeed, it is
generally necessary to append to quantum theory an additional
postulate to the effect that only as a consequence of another kind
of a sudden perturbation, namely, an operation of measurement,
will the wave function of the system be able to `jump' into one or
another of the eigenstates of the new Hamiltonian.

In contrast, by use of the stochastic extension to the
Schr\"odinger equation that we discuss in \S~3, it is possible to
model the dynamics of the wave function in such a manner that,
after the system is perturbed, the wave function spontaneously
relaxes to one of the eigenstates of the new Hamiltonian. It is a
remarkable fact that the probability laws thus arising from the
stochastic dynamics give rise to statistical predictions that are
essentially in agreement with the standard predictions of quantum
theory.

The purpose of this paper is to present a detailed analysis of the
phenomenon of stochastic relaxation in the simple case of a
particle in a potential well when the width of the well is
instantaneously enlarged. The structure of the paper is as
follows. In \S~2 we review the basic setup for a free particle in
a potential well when the well is subjected to a sudden expansion.
We argue that standard quantum mechanics does not give a
completely satisfactory account of the matter. In \S~3 we review
the formalism of the standard energy-based stochastic extension of
the Schr\"odinger equation, and propose the use of stochastic
relaxation as a basis for the description of the dynamics of a
quantum system following a perturbation. In \S\S~4--5 we construct
the solution to the stochastic system and describe its properties,
and in \S\S~6--7 we indicate how the solution can be used for the
efficient construction of simulations. More precisely, we show
that, according to the stochastic evolutionary law, the system
energy fluctuates randomly and eventually relaxes to one of the
eigenvalues of the new Hamiltonian. Graphic illustrations are
provided for the behaviour of the energy, as well as the
probability density of the location of the particle in the well.
In \S~8 we present an analysis of the timescale associated with
the eventuality of relaxation. This analysis then forms the basis
of a discussion of a stochastic version of the quantum adiabatic
theorem, presented in \S\S~9--10.

\section{Free expansion in a potential well}

We analyse here the sudden expansion of a one-dimensional
potential well in which a particle of mass $\mu$ is trapped.
Before the expansion, the width of the well is $L$. The energy
spectrum of the particle is given by
\begin{eqnarray}
\epsilon_n = \frac{\pi^2\hbar^2 n^2}{2\mu L^2}, \label{eq:2}
\end{eqnarray}
where $n=1,2,\ldots,\infty$, for which the corresponding
eigenfunctions are
\begin{eqnarray}
\phi_n(x) = \sqrt{\frac{2}{L}}\sin\left( \frac{n\pi}{L}x\right),
\quad\quad 0\leq x\leq L . \label{eq:3}
\end{eqnarray}
We assume that initially the particle is in one of the energy
eigenstates associated with the potential.

Let us suppose that at $t=0$ the Hamiltonian ${\hat h}$ is changed
in such a way that the width of the potential is increased from
$L$ to $\alpha L$, where $\alpha\geq1$. Then for any $t>0$ the
wave function of the system can be expressed as a superposition of
the normalised stationary states of the new Hamiltonian ${\hat
H}$. These are given by
\begin{eqnarray}
\chi_n(x) = \sqrt{\frac{2}{\alpha L}}\sin\left( \frac{n\pi}{\alpha
L}x\right), \quad\quad 0\leq x\leq \alpha L , \label{eq:4}
\end{eqnarray}
for which the associated eigenvalues are
\begin{eqnarray}
E_n = \frac{\pi^2\hbar^2 n^2}{2\mu \alpha^2L^2}, \label{eq:4.1}
\end{eqnarray}
where $n=1,\ldots,\infty$. It follows according to the unitary
dynamics of the Schr\"odinger equation that after the expansion
has taken place the system will be in an indefinite state of
energy, and will remain so.

If the particle is initially in the $n$-th eigenstate $\phi_n(x)$
with energy $\epsilon_n$, then the probability $\pi_{nm}$ that,
after the expansion has taken place, the particle will be found in
the $m$-th eigenstate $\chi_m(x)$ of the new Hamiltonian, with
energy $E_m$, is
\begin{eqnarray}
\pi_{nm} = \left( \int_0^L \phi_n(x) \chi_m(x) {\rm d}x\right)^2.
\end{eqnarray}
A short calculation shows that
\begin{eqnarray}
\pi_{nm} = \frac{4\alpha^3 n^2 }{\pi^2(m^2-\alpha^2 n^2)^2}
\sin^2\left(\frac{\pi m} {\alpha}\right) . \label{eq:5}
\end{eqnarray}
Clearly, if such a change of state occurs, then energy is not
strictly conserved. The conservation of energy is maintained in
expectation, however. Indeed, we have the identity
\begin{eqnarray}
\sum_{m=1}^\infty \pi_{nm} E_m = \epsilon_n ,\label{eq:6}
\end{eqnarray}
which is valid for all $n$ and for all $\alpha$, by virtue of
which we can confirm that the expectation value of the Hamiltonian
is a constant of the motion (Bender, Brody $\&$ Meister 1999).

To understand the energy conservation law intuitively in this
example, we can regard the potential walls as forming the `piston'
of a one-dimensional cylinder. Suppose we consider an ensemble
consisting of a large number of independent identical particles in
the cylinder. Then when the piston is instantaneously moved
outwards, the system undergoes a free expansion. During such a
process, the particles do no mechanical work on the piston. As a
result, there is no net flow of energy going out of the cylinder,
and thus the energy is conserved. The `energy' that is conserved
in this case is the ensemble energy, i.e. the product of total
number of particles and the expectation value of the energy of an
individual particle.

The foregoing formulae follow directly from the basic principles
of quantum theory. In particular, the standard interpretation of
quantum theory maintains that after the expansion the system will
remain in an indefinite state of energy, until the energy is
measured. However, there are many situations in which it is
natural to presume that after the passage of some time the system
spontaneously relaxes into one or another of the eigenstates of
the new Hamiltonian, irrespective of whether a measurement is
made. Many natural phenomena are of this character: after
perturbation, there follows relaxation. Quantum theory, as such,
does not account for this satisfactorily. There is arguably an
implicit assumption that, in the absence of specific acts of
measurement, quantum systems of any significant size or complexity
will settle into a stable eigenstate, typically an eigenstate of
energy.

In what follows we take the point of view that, after
perturbation, the system does eventually evolve spontaneously into
one of the stable eigenstates. Quantum theory in itself offers no
clue as regards the precise mechanism according to which the
system relaxes to the new stationary state. These questions can,
nevertheless, be addressed in a very satisfactory way by use of
the dynamics of the stochastic extension of the Schr\"odinger
equation, as we shall demonstrate.

\section{Energy-based stochastic dynamics}

In what follows we shall suppose that the dynamical mechanism
governing the relaxation of the quantum system is determined by
the standard energy-based stochastic extension of the
Schr\"odinger equation. This is given by the following stochastic
differential equation of the Ito type:
\begin{eqnarray}
{\rm d}\psi_t(x) = -{\rm i}{\hat H}\psi_t(x){\rm d}t - \octa
\sigma^2 ({\hat H}-H_t)^2 \psi_t(x) {\rm d}t + \half \sigma ({\hat
H}-H_t)\psi_t(x) {\rm d}W_t \label{eq:3.1}
\end{eqnarray}
for which it is assumed that there is a prescribed initial wave
function $\psi_0(x)$. Here $W_t$ denotes a standard Wiener
process, and
\begin{eqnarray}
H_t = \frac{\int \psi_t^*(x) {\hat H} \psi_t(x) {\rm d}x} {\int
\psi_t^*(x) \psi_t(x) {\rm d}x}  \label{eq:1.2}
\end{eqnarray}
is the random process corresponding to the expectation value of
the Hamiltonian operator ${\hat H}$ in the random state
$\psi_t(x)$. The volatility parameter $\sigma$ appearing in
(\ref{eq:3.1}) has the units
\begin{eqnarray}
[\sigma] = [{\rm Energy}]^{-1}[{\rm Time}]^{-1/2}. \label{eq:1.3}
\end{eqnarray}
Dynamical equations of the type (\ref{eq:3.1}) for the evolution
of the wave function and various generalisations thereof, were
introduced originally as simple models to characterise the
collapse of the wave function when a measurement is carried out on
a system (Gisin 1984, 1989, Ghirardi, {\em et al} 1986, 1990,
Diosi 1988; see, e.g., Percival 1998, Pearle 2000 and Bassi $\&$
Ghirardi 2002 and references cited therein for a more
comprehensive account of the relevant literature). The idea that
wave function should proceed to energy eigenstates was proposed by
Bedford and Wang (1975, 1977). The specific energy-based form of
the dynamics (\ref{eq:3.1}), which has been studied by Gisin 1989,
Percival 1994, 1995, Hughston 1996, Adler $\&$ Horowitz 2000,
Adler 2002, and Brody $\&$ Hughston 2002, amongst others, is the
most parsimonious of these state reduction models and in many
respects perhaps the most attractive as the basis for a
fundamental model. In this paper we carry the physical application
of the stochastic theory a step further and propose the use of
(\ref{eq:3.1}) as an elementary model for characterising the
relaxation of the state of a quantum system when its Hamiltonian
has been perturbed. We shall leave open here the question of
whether the volatility parameter $\sigma$ governing the timescale
of relaxation is phenomenological in character, i.e. varying
according to the structure of the system, or universal, e.g.,
Planckian. Before deriving the solution to (\ref{eq:3.1}) and
applying the results to perturbation theory, we shall briefly
sketch some of the basic mathematical and physical properties
associated with the dynamics (\ref{eq:3.1}). For further details,
see Adler {\em et al}. (2001) and references cited therein.

Let us first note that the coefficient of the Brownian motion in
the third term of the right-hand side of (\ref{eq:3.1}) is given
by the difference of the Hamiltonian operator and its expectation,
acting on the wave function. Thus if the system enters into an
eigenstate of the Hamiltonian, this coefficient becomes zero, and
the random fluctuations generated by the process $W_t$ make no
further contribution to the dynamics of $\psi_t(x)$. This property
likewise applies to the second term, which together with the third
term `drives' the system into a state of lower energy uncertainty.
Starting from an arbitrary initial state, the system is randomly
driven into states with lower energy variance, until it finally
reaches an eigenstate of the Hamiltonian, in which the variance
vanishes.

A process for which both of the coefficients of the ${\rm d}t$
term and the ${\rm d}W_t$ term in the associated stochastic
differential equation are smooth functions of the process
$\psi_t(x)$ itself is called a {\sl diffusion}. If the number of
Brownian motions is smaller than the dimensionality of the process
$\psi_t(x)$, then the process is said to be a {\sl degenerate}
diffusion. In the present consideration, while the wave function
$\psi_t(x)$ is an element of an infinite dimensional Hilbert
space, there is only a single Brownian motion driving the system,
and thus the dynamical equation (\ref{eq:3.1}) represents a highly
degenerate diffusion. A typical feature of a degenerate diffusion,
as opposed to a generic diffusion, is that it may have a
`focusing' effect. Indeed, in the case of (\ref{eq:3.1}) the wave
function is focussed to one of the energy eigenstates, as a
consequence of the dynamics.

Given the dynamical equation (\ref{eq:3.1}) for the wave function
and the corresponding process (\ref{eq:1.2}) for the expected
energy, we can determine the stochastic equation satisfied by
$H_t$. Specifically, this is given by
\begin{eqnarray}
{\rm d}H_t = \sigma V_t {\rm d}W_t , \label{eq:2.2}
\end{eqnarray}
where
\begin{eqnarray}
V_t = \frac{\langle{\psi}_t|({\hat H}-H_t)^2|\psi_t\rangle}
{\langle{\psi}_t|\psi_t\rangle}  \label{eq:1.5}
\end{eqnarray}
is the process associated with the variance of the energy. The
variance process satisfies
\begin{eqnarray}
{\rm d}V_t = -\sigma^2 V_t^2 {\rm d}t + \sigma \beta_t {\rm d}W_t
, \label{eq:var}
\end{eqnarray}
where
\begin{eqnarray}
\beta_t = \frac{\langle{\psi}_t|({\hat H}-H_t)^3|\psi_t\rangle}
{\langle{\psi}_t|\psi_t\rangle}  \label{eq:2.7}
\end{eqnarray}
is the third central moment of the energy. As a consequence,
integrating (\ref{eq:2.2}) and (\ref{eq:var}) we find that the
energy process can be expressed in the form
\begin{eqnarray}
H_t = H_0 + \sigma \int_0^t V_s {\rm d}W_s  \label{eq:2.5}
\end{eqnarray}
and the variance process can be written as
\begin{eqnarray}
V_t = V_0 - \sigma^2 \int_0^t V_s^2 {\rm d}s + \sigma \int_0^t
\beta_s {\rm d}W_s . \label{eq:2.6}
\end{eqnarray}

Owing to elementary properties of the stochastic integrals
(\ref{eq:2.5}) and (\ref{eq:2.6}) one can determine the relations
\begin{eqnarray}
{\mathbb E}\left[ H_u|\{H_s\}_{0\leq s\leq t}\right] = H_t
\label{eq:2.55}
\end{eqnarray}
and
\begin{eqnarray}
{\mathbb E}\left[ V_u|\{V_s\}_{0\leq s\leq t}\right] \leq V_t
\label{eq:2.65}
\end{eqnarray}
for $t\leq u$. Here ${\mathbb E}[\cdots|\{X_s\}_{0\leq s\leq t}]$
denotes the conditional expectation given the history of the
process $X_s$ from time $0$ up to time $t$. Therefore, the
conditional expectation of the energy process at any time $u\geq
t$, given its history up to time $t$, is given by its value at
time $t$. We thus say that $H_t$ satisfies the {\sl martingale}
condition. It follows that $H_t$ is on average conserved, whereas
the variance $V_t$ tends, on average, to decrease, corresponding
to the spontaneous reduction of the system to an eigenstate. The
reductive character of the dynamics (\ref{eq:3.1}) is indicated by
the fact that $V_t$ satisfies the {\sl supermartingale} condition.

\section{Analytic solution to the stochastic equation}

Despite its nonlinearity, the dynamical equation (\ref{eq:3.1})
can be solved exactly to yield an analytic solution that fully
characterises the dynamical state of the system at any time in
terms of a pair of state variables (Brody $\&$ Hughston 2002). The
method of obtaining the solution, which is of some interest in its
own right, makes use of the classical techniques of nonlinear
filtering theory. More precisely, we consider a process of the
form
\begin{eqnarray}
\xi_t = \sigma t H + B_t , \label{eq:7}
\end{eqnarray}
where $B_t$ is a Brownian motion, and $\sigma$ is the parameter
that ultimately determines the characteristic relaxation
timescale. The random variable $H$ takes on the value $E_m$ with
probability $\pi_{m}$, where $\pi_m$ is taken to be the transition
probability from the initial quantum state to the energy
eigenstate with energy $E_m$. Intuitively, one can think of the
process $\xi_t$ as representing the value of a phase, scaled by
the constant $\sigma$, together with a random noise term, whose
strength, relative to the variable $H$, decreases
inverse-proportionally in time.

The random variable $H$ is to be thought of as representing the
value of the energy to which the system ultimately relaxes after
the passage of sufficient time. Given the trajectory of the
process $\xi_t$ up to time $t$, we would like to determine the
best estimate for the value of $H$. Because the standard error in
the value of $B_t$ grows only like the square-root of time, it
follows that as time passes the true value of $H$ is gradually
revealed. The estimate will be denoted $H_t$, which is obtained by
taking the conditional expectation of the random variable $H$,
given the history of $\xi_t$ up to that time. Because $\xi_t$ is a
Markov process, this implies that $H_t$ is the expectation of $H$
conditional on the value $\xi_t$, that is,
\begin{eqnarray}
H_t = {\mathbb E}\left[ H|\xi_t\right] . \label{eq:8}
\end{eqnarray}
The argument that $H_t$ is the best estimate for $H$ given the
history of $\xi_t$ is as follows. Suppose $Y_t$ is any process
that at time $t$ can be expressed as a functional of the history
of $\xi_s$ for $0\leq s\leq t$. Then the choice of $Y_t$ that
minimises the expected mean square error ${\mathbb E}[(H-Y_t)^2|
\{\xi_s \}_{0\leq s\leq t}]$ given the history of $\xi_t$ is the
process $H_t$ defined by (\ref{eq:8}). This can be deduced by a
straightforward variational argument.

Because $H_t$ is given, for each value of $t$, by a conditional
expectation with respect to the random variable $\xi_t$, it
follows that $H_t$ can be expressed as a function of $\xi_t$. In
order to determine the conditional expectation (\ref{eq:8}) more
explicitly, we require the conditional probability ${\mathbb
P}(H=E_m|\xi_t)$ for the random variable $H$. By use of the Bayes
law for conditional probability, we find that this is given by
\begin{eqnarray}
{\mathbb P}(H=E_m|\xi_t) = \frac{\pi_{m}\rho (\xi_t|H=E_m)}
{\sum_{n=1}^\infty \pi_{n} \rho(\xi_t|H=E_n)}. \label{eq:9}
\end{eqnarray}
Here $\rho(\xi_t|H=E_m)$ denotes the conditional density function
for the continuous random variable $\xi_t$ given that $H=E_m$. In
deriving (\ref{eq:9}) we make use of the fact that the {\sl
unconditional} probability ${\mathbb P}(H=E_m)$ is just $\pi_{m}$
for the given initial state. We have also used the relation
\begin{eqnarray}
\rho(\xi_t) = \sum_{n=1}^\infty \pi_{n} \rho(\xi_t|H=E_n).
\label{eq:9.5}
\end{eqnarray}
Since $B_t$ is a Brownian motion, it is by definition normally
distributed with mean zero and variance $t$. Thus the conditional
probability density for $\xi_t$ is
\begin{eqnarray}
\rho(\xi_t|H=E_m) = \frac{1}{\sqrt{2\pi t}} \exp\left(
-\frac{1}{2t} (\xi_t-\sigma E_m t)^2\right). \label{eq:10}
\end{eqnarray}
Inserting this expression into (\ref{eq:9}), we deduce that
\begin{eqnarray}
{\mathbb P}(H=E_m|\xi_t) &=& \frac{\pi_{m} \exp\left( \sigma E_m
\xi_t -\half \sigma^2 E_m^2 t\right)}{\sum_{n=1}^\infty \pi_{n}
\exp\left( \sigma E_n \xi_t-\half \sigma^2 E_n^2 t\right)}.
\label{eq:11}
\end{eqnarray}
Since the energy process is given by the expectation
\begin{eqnarray}
H_t = \sum_{m=1}^\infty E_m {\mathbb P}(H=E_m|\xi_t),
\end{eqnarray}
it follows that
\begin{eqnarray}
H_t = \frac{\sum_{m=1}^\infty \pi_{m} E_m \exp\left( \sigma E_m
\xi_t-\half \sigma^2 E_m^2 t\right)}{\sum_{m=1}^\infty \pi_{m}
\exp\left( \sigma E_m \xi_t-\half \sigma^2 E_m^2 t\right)}.
\label{eq:12}
\end{eqnarray}
We note that in (\ref{eq:12}) the process $H_t$ for the
conditional expectation of the random variable $H$ is expressed in
terms of a function of $t$ and $\xi_t$. As a consequence, the
dynamics of the process $\xi_t$ can be expressed as a diffusion
equation of the form
\begin{eqnarray}
{\rm d}\xi_t = \sigma H_t {\rm d}t + {\rm d}W_t, \label{eq:13}
\end{eqnarray}
where $W_t$ is a standard Brownian motion with respect to the
filtration generated by the history of $H_t$. The existence of a
Brownian motion $W_t$ satisfying (\ref{eq:13}) follows as a
consequence of well-established line of argument in nonlinear
filtering theory (see Liptser $\&$ Shiryaev 1974), the details of
which in the present context are set out in Brody $\&$ Hughston
(2002).

\section{Stochastic relaxation}

In order to see directly the relaxation of the Hamiltonian process
(\ref{eq:12}) to one of the energy eigenvalues, we can argue as
follows.

Let us suppose that the random variable $H$ happens to take the
value $E_j$. By definition such an event occurs with probability
$\pi_{j}$. More precisely, we condition on the outcome of the
event $H=E_j$ and analyse the evolution of process (\ref{eq:12}).
Then, writing $\xi_t=\sigma E_j t+B_t$, we obtain, for the
corresponding realisation of $H_t$, the expression
\begin{eqnarray}
H_t^j = \frac{\pi_{j} E_j + \sum_{m\neq j}^\infty \pi_{m} E_m
\exp\left( \sigma (E_m-E_j) B_t-\half \sigma^2 (E_m-E_j)^2
t\right)}{\pi_{j}+\sum_{m\neq j}^\infty \pi_{m} \exp\left( \sigma
(E_m-E_j) B_t-\half \sigma^2 (E_m-E_j)^2 t\right)} \label{eq:14}
\end{eqnarray}
where the superscript $j$ in $H_t^j$ indicates the conditioning on
the event $\{H=E_j\}$.

Because the exponential terms appearing in both denominator and
numerator have the property that, as $t\rightarrow\infty$, the
probability that these terms remain positive approaches zero, it
follows that the energy process $H_t^j$ converges asymptotically
to the designated eigenvalue $E_j$.

Indeed for any normally distributed random variable $B_t$ with
mean zero and variance $t$ it is an elementary fact that if
$\nu\neq0$ then
\begin{eqnarray}
\lim_{t\to\infty} {\mathbb P}\left( \exp\left( \nu B_t-\half \nu^2
t\right) > x\right) = 0
\end{eqnarray}
for any given $x>0$.

The exact, closed-form solution (\ref{eq:14}) shows how the
expectation value of the system Hamiltonian evolves in time. In
particular, it shows how the system organises itself spontaneously
to move into a stable, stationary state of the new Hamiltonian.
Since formula (\ref{eq:14}) is expressed in terms of a simple
analytic function of the Brownian motion $B_t$ and time $t$, there
arises the possibility of efficiently simulating the evolution of
the process $H_t$.

\section{Wave function dynamics}

We return now to the analysis of the dynamics of the state
$\psi_t(x)$ of a particle trapped in a potential well following a
sudden expansion of the width of the well. In this case the
solution to the stochastic differential equation (\ref{eq:3.1})
can be expressed in the form
\begin{eqnarray}
\psi_t(x) = \frac{\sum_{m=1}^\infty \pi_{m}^{1/2} \exp\left( -{\rm
i}E_m t+\frac{1}{2}\sigma E_m\xi_t - \frac{1}{4}\sigma^2 E_m^2
t\right)\chi_m(x)}{\left( \sum_{m=1}^\infty \pi_{m} \exp\left(
\sigma E_m\xi_t - \frac{1}{2}\sigma^2 E_m^2 t
\right)\right)^{1/2}}. \label{eq:16}
\end{eqnarray}
Here $\chi_m(x)$ denotes the normalised eigenfunction of the
Hamiltonian $\hat{H}$ with energy $E_m$, and the choice of the
initial wave function $\psi_0(x)$ is implicit in the probability
$\pi_{m}$:
\begin{eqnarray}
\pi_{nm} = \left( \int_0^{L} \psi_0(x) \chi_m(x) {\rm d}x
\right)^2.
\end{eqnarray}
The solution (\ref{eq:16}) can be verified by taking the
stochastic differential of the right hand side of (\ref{eq:16})
and using the Ito rules, and then making the substitution
(\ref{eq:13}).

The convergence of the infinite sum in the numerator and the
denominator of (\ref{eq:16}) for finite $t$ may not be immediately
evident. In order to check that these summations converge, we
substitute a particular realisation of the path for $\xi_t$, say,
$\xi_t=\sigma E_j t+B_t$, when the random variable $H$ happens to
take the value $E_j$. Then on account of the fact that $E_m\propto
m^2$, we find that the summands decay, for each fixed $j$, like
$\sim\exp(-m^4)$, and convergence is ensured.

Let us now consider the random dynamics of the probability density
function
\begin{eqnarray}
\rho_t(x)= \psi_t^*(x) \psi_t(x)
\end{eqnarray}
for finding the particle at the location $x$ in the interval
$[0,\alpha L]$. We would like to obtain the probability
distribution for the particle in the case where the final state of
the system is given by the eigenstate $\chi_j(x)$ for some given
value of $j$. The probability for this particular realisation to
occur is given by $\pi_{j}$. By rearrangement of terms, and
writing
\begin{eqnarray}
\omega_{mj}=E_m-E_j
\end{eqnarray}
for the energy-level difference, we obtain
\begin{eqnarray}
\rho_{t}^j(x) = \frac{\left| \sum_m \pi_{m}^{1/2} \exp\left( -{\rm
i}\omega_{mj} t + \frac{1}{2}\sigma \omega_{mj}B_t -
\frac{1}{4}\sigma^2 \omega_{mj}^2 t\right)\chi_m(x)\right|^2}{
\sum_m \pi_{m} \exp\left( \sigma \omega_{mj}B_t -
\frac{1}{2}\sigma^2 \omega_{mj}^2 t \right)} \label{eq:17}
\end{eqnarray}
for the probability density. Thus $\rho_t(x)$ is a measure-valued
process, i.e. at each given time $t$ it is a smooth density
function over $[0,\alpha L]$, but the form of the function evolves
randomly in time until it relaxes to the final distribution
$\chi_j^2(x)$.

\section{Simulation of the energy and the probability density}

\begin{figure}
\label{fig1}
 {\centerline{\psfig{file=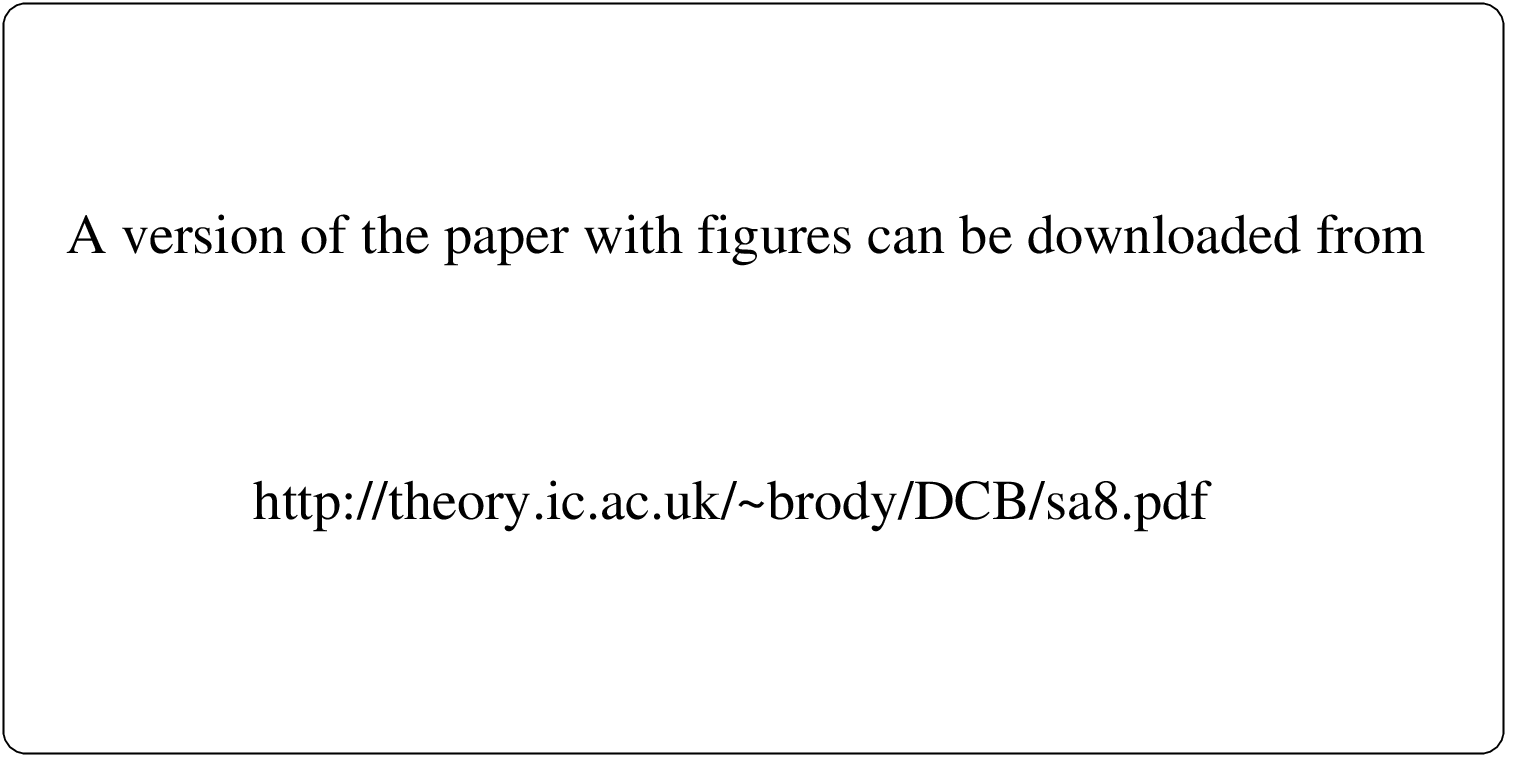,width=12cm,angle=270}}}
\vspace{0cm} \caption{Several realisations of the process $H_t$
for the expectation value of the Hamiltonian are illustrated. The
initial condition is chosen to be $\psi_0(x)=\phi_1(x)$ and the
final conditions are given by $\psi_\infty(x)=\chi_1(x),\chi_2(x),
\chi_3(x),\chi_4(x),\chi_5(x),\chi_6(x)$, respectively. The
horizontal lines indicate the energy levels $E_j$, where we have
set $\alpha=2.5$. For convenience the vertical axis is expressed
in units of the characteristic energy $\varepsilon=\hbar^2/2\mu
L^2$, and the horizontal axis is expressed in units of the
corresponding characteristic time interval
$1/\sigma^2\varepsilon^2$. When $j=5$, we have $\pi_{5}=0$. As a
result, if $E_j=E_5$ is chosen for the value of the random
variable, the energy does not reduce to that eigenvalue.}
\end{figure}

With these formulae at hand, we now consider the simulation of the
random trajectory for the expectation value of the Hamiltonian
governed by the dynamics (\ref{eq:12}). We also consider the
simulation of the corresponding probability density function
(\ref{eq:17}) for the position of the particle in the potential
well. The quantum system is defined by specifying the value of the
mass $\mu$ of the particle, the width $L$ of the well, the
volatility parameter $\sigma$ governing the stochastic dynamics,
and the well expansion factor $\alpha$. For simplicity, we shall
examine the case where the system is initially in its ground state
$\phi_1(x)$, although any other initial condition, including mixed
initial states, can be treated analogously. Then the energy
process (\ref{eq:14}), corresponding to the particular realisation
$\psi_\infty(x)=\chi_j(x)$ of the terminal state, can be written
in the form
\begin{eqnarray}
H_t^j = \frac{ \sum_{m} \pi_{m} E_m \exp\left( \sigma \omega_{mj}
B_t-\half \sigma^2 \omega_{mj}^2 t\right)}{\sum_{m} \pi_{m}
\exp\left( \sigma \omega_{mj}B_t-\half \sigma^2 \omega_{mj}^2
t\right)}. \label{eq:18}
\end{eqnarray}
The associated probability density for the position of the
particle in the interval $[0,\alpha L]$ is given by
\begin{eqnarray}
\rho_{t}^j(x) &=& \frac{\left( \sum_m \pi_{m}^{1/2} \exp\left(
\frac{1}{2}\sigma \omega_{mj}B_t - \frac{1}{4}\sigma^2
\omega_{mj}^2 t\right) \chi_m(x) \cos(\omega_{mj}t)\right)^2 }{
\sum_m \pi_{m} \exp\left( \sigma \omega_{mj}B_t -
\frac{1}{2}\sigma^2 \omega_{mj}^2 t \right)} \nonumber \\ & & +
\frac{\left( \sum_m \pi_{m}^{1/2} \exp\left( \frac{1}{2}\sigma
\omega_{mj}B_t - \frac{1}{4}\sigma^2 \omega_{mj}^2 t\right)
\chi_m(x) \sin(\omega_{mj}t)\right)^2 }{ \sum_m \pi_{m} \exp\left(
\sigma \omega_{mj}B_t - \frac{1}{2}\sigma^2 \omega_{mj}^2 t
\right)}. \label{eq:19}
\end{eqnarray}

\begin{figure}
\label{fig2}
 {\centerline{\psfig{file=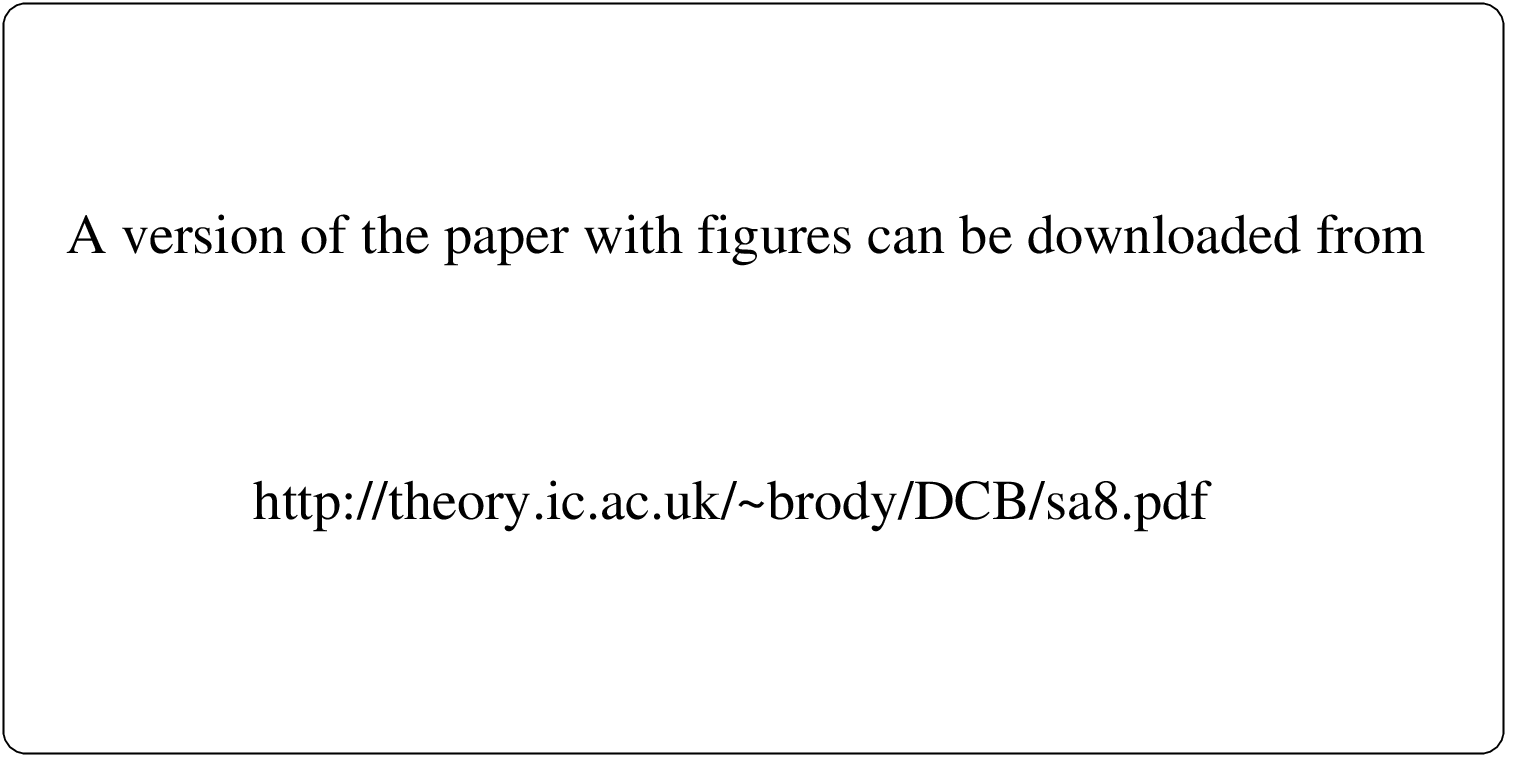,width=12cm,angle=270}}}
\vspace{0cm} \caption{Several realisations of the probability
distributions $\rho_t(x)$ for the particle in the well are
illustrated. The initial and final conditions are
$\psi_0(x)=\phi_1(x)$ and $\psi_{\infty}(x)=\chi_2(x)$,
respectively, and we set $\alpha=2.5$. The $x$-axis is measured in
units of $L$, and the time axis is measured in units of
$4\mu^2L^4/\hbar^4\sigma^2$. Most realisations have the property
that the initial state swiftly relaxes into the terminal state. }
\end{figure}

The process $H_t^j$ defined by (\ref{eq:18}) can be thought of as
a kind of `Brownian bridge' that interpolates between the two
energy levels $\epsilon_1$ and $E_j$. That is to say, initially
the system has energy $\epsilon_1$, and then it progresses along a
random trajectory to reach the designated terminal level $E_j$. A
similar remark applies to the density process $\rho_{t}^j(x)$ of
(\ref{eq:19}), which interpolates between the two functions
$\phi_1^2(x)$ and $\chi_j^2(x)$. It should be evident that, unlike
$H_t$, the process $H_t^j$ does not conserve energy.

To obtain a better feeling for the dynamics of $H_t^j$ it will be
useful to examine the associated stochastic differential equation.
In particular, if we take the stochastic differential of
(\ref{eq:18}), then after a rearrangement of terms we obtain
\begin{eqnarray}
{\rm d}H_t^j = \sigma^2 V_t^j \left( E_j-H_t^j\right) {\rm d}t +
\sigma V_t^j {\rm d}B_t , \label{eq:19.5}
\end{eqnarray}
where the nonnegative process $V_t^j$ is given by
\begin{eqnarray}
V_t^j = \frac{\sum_{m}\pi_{m} (E_m-H_t^j)^2 \exp\left(\sigma
\omega_{mj}B_t -\half \sigma^2 \omega_{mj}^2 t\right)}{\sum_{k}
\pi_{m} \exp\left( \sigma \omega_{mj}B_t -\half \sigma^2
\omega_{mj}^2 t\right)}. \label{eq:19.6}
\end{eqnarray}
We learn from (\ref{eq:19.5}) that $H_t^j$ is a {\sl
mean-reverting} process with mean level $E_j$ and reversion rate
$\sigma^2 V_t^j$. In fact, one can integrate (\ref{eq:19.5}) by
means of the standard technique used in the case of the classical
Ornstein-Uhlenbeck process (see, e.g., Doob 1942) to obtain
\begin{eqnarray}
H_t^j = E_j + \left( H_0^j-E_j\right) {\rm e}^{-\sigma^2 \int_0^t
V_s^j{\rm d}s}+\sigma\int_0^t{\rm e}^{-\sigma^2\int_u^t V^j_s{\rm
d}s} V_u^j {\rm d}B_u , \label{eq:6.5}
\end{eqnarray}
which shows clearly how the information of the initial condition
is damped away at the rate $\sigma^2 {\bar V}_t^j$, where
\begin{eqnarray}
{\bar V}_t^j = \frac{1}{t} \int_0^t V_s^j {\rm d}s.
\end{eqnarray}
Alternatively, we can write (\ref{eq:6.5}) in the form
\begin{eqnarray}
H_t^j - E_j = \left( H_0^j-E_j + \sigma\int_0^t{\rm
e}^{\sigma^2\int_0^u V_s^j{\rm d}s} V_u^j {\rm d}B_u \right) {\rm
e}^{-\sigma^2 \int_0^t V_s^j{\rm d}s} ,
\end{eqnarray}
which expresses the difference $H_t^j-E_j$ between $H_t^j$ and its
terminal value as the product of a martingale and a positive
decreasing process.

With expressions (\ref{eq:18}) and (\ref{eq:19}) at hand we
proceed to simulate some realisations of these processes. For
convenience we take advantage of the fact that there is a natural
energy unit $\varepsilon$ determined by the problem, which is
given by
\begin{eqnarray}
\varepsilon = \frac{\hbar^2}{2\mu L^2}.
\end{eqnarray}
Thus, when we plot figures we can express energies in units of
$\varepsilon$, and we can express times in units of
\begin{eqnarray}
\frac{1}{\sigma^2\varepsilon^2} =
\frac{4\mu^2L^4}{\hbar^4\sigma^2} .
\end{eqnarray}
In the analysis that follows, energy and time will be expressed in
these units.

In Figure~1 the energy process $H_t^j$ is shown for several values
of $j$ ranging from 1 to 6, where initially the system is in the
ground state of the old potential, with energy $\epsilon_1=\pi^2$.
In these examples the width of the potential is expanded at $t=0$
by a factor of $\alpha=2.5$. The most likely transition to occur,
given the initial state $\phi_1(x)$, when the width is expanded by
a factor of 2.5, is the first excited state $\chi_{2}(x)$. In this
example it follows as a consequence of (\ref{eq:5}) that
$\pi_{15}=0$, and hence the transition into the fifth energy level
does not occur. Instead, there is a transition to the fourth
level.

\begin{figure}
\label{fig3}
 {\centerline{\psfig{file=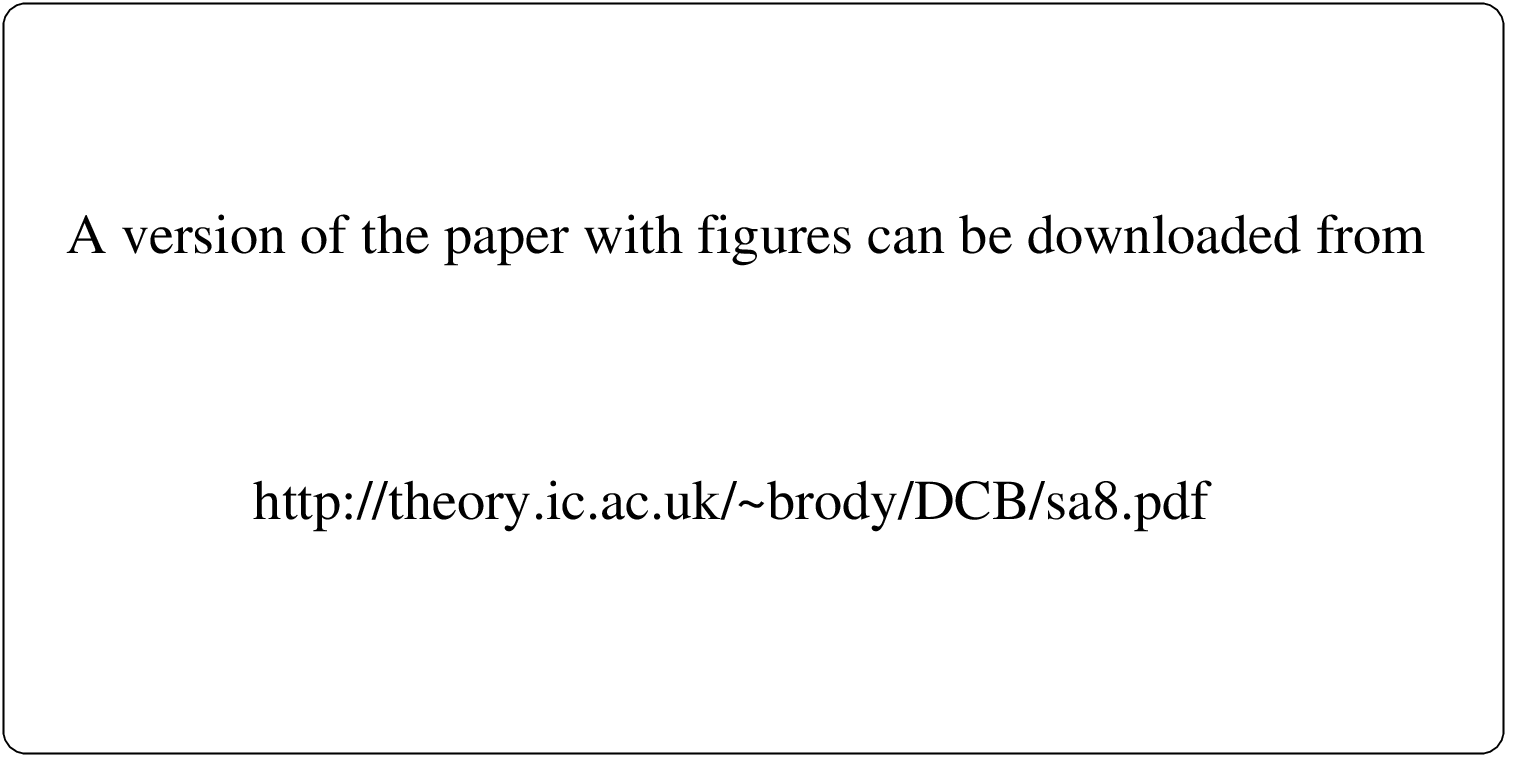,width=12cm,angle=270}}}
\vspace{0cm} \caption{The expectation of $H_t^j$ is shown for a
variety of realisations $j=1,2,3,4,5,6$, when the initial state is
$\phi_1(x)$. The curious behaviour for $j=5$ arises in this
example because $\pi_5=0$, and the state never collapses to
$\chi_5(x)$. Instead, the state reduced to the eigenstate having
the energy closest to $E_5$, which in the present case is the
fourth eigenstate.}
\end{figure}

Several realisations of the corresponding probability distribution
for the particle are shown in Figure~2, where the terminal wave
function is the first excited state $\chi_2(x)$. The probability
of this event to be realised is as large as $\sim 0.43$. In most
examples the initial state swiftly changes into the terminal
state, although some interesting behaviour can occasionally be
observed when the value of the Brownian motion $B_t$ grows large.

In Figure~3 the numerical average of the energy process,
corresponding to the expectation of $H_t^j$, is shown for a range
of terminal states in the case where the expansion factor is
$\alpha=2.5$. Each graph represents the average of one thousand
simulations for each energy level. The behaviour observed in the
fifth energy level reflects the fact that, in this case, the
terminal state is given by the fourth eigenstate $\chi_4(x)$. This
is because $E_4$ is the closest energy eigenvalue to $E_5$ in the
present example. To see this, we note that, if we write
$\xi=\exp(\sigma\omega_{45}B_t - \frac{1}{2}
\sigma^2\omega_{45}^2t)$ then as a consequence of (\ref{eq:18}) it
follows that $H_t^5$ can be written in the form
\begin{eqnarray}
H_t^5 = \frac{ \xi\left( \pi_4 E_4 + \sum_{m}' \pi_{m} E_m
\exp\left( \sigma (\omega_{m5}-\omega_{45}) B_t-\half \sigma^2
(\omega_{m5}^2-\omega_{45}^2) t\right)\right)}{\xi \left( \pi_4 +
\sum_{m}' \pi_{m} \exp\left( \sigma (\omega_{m5}-\omega_{45})
B_t-\half \sigma^2 (\omega_{m5}^2-\omega_{45}^2) t\right)\right)}
\label{eq:18-5}
\end{eqnarray}
where $\sum'_{m}=\sum_{m\neq4}$. Thus, the $\xi$-dependence
cancels, and because $\omega_{m5}^2- \omega_{45}^2>0$ for all
$m\neq4$, we see that as $t\to\infty$ the exponents in the
denominator and the numerator go to zero and we are left with the
leading term $E_4$.

In general, if the terminal value for the energy is chosen to be
$E_j$ when the initial state $\psi_0(x)$ is orthogonal to
$\chi_j(x)$, i.e. $\pi_j=0$, then the system necessarily relaxes
into the eigenstate whose eigenvalue is the closest to $E_j$
(S.~L.~Adler, private communication).

In Figure~4 we sketch the dynamics of the ensemble average for the
probability density function for the location of the particle, in
the case where the terminal state is the first excited state
$\chi_2(x)$. This result corresponds to the average of a thousand
examples of the type presented in Figure~2.

\section{Relaxation timescale}

One of the advantages of the simulation methodology is that it
opens up the possibility of a direct analysis of the timescale
$\tau_R$ over which relaxation typically occurs. Let us define the
process $M_{mj}$ by
\begin{eqnarray}
M_{mj} = \exp\left( \sigma \omega_{mj} B_t - \half \sigma^2
\omega_{mj}^2 t \right) , \label{eq:20}
\end{eqnarray}
where for simplicity of notation we suppress the time dependence
of $M_{mj}(t)$. The energy process (\ref{eq:18}) conditional on
the outcome $E_j$ can then be written in the form
\begin{eqnarray}
H_t^j = \frac{\pi_j E_j + \sum_{m\neq j}^\infty \pi_m E_m M_{mj}}
{\pi_j + \sum_{m\neq j}^\infty \pi_m M_{mj}}, \label{eq:21}
\end{eqnarray}
since $M_{jj}=1$. Clearly, if $M_{mj}$ is sufficiently small, then
$H_t^j$ will approach its terminal value $E_j$ and the system will
have relaxed. This occurs when $t>\tau_R$.

\begin{figure}
\label{fig4}
 {\centerline{\psfig{file=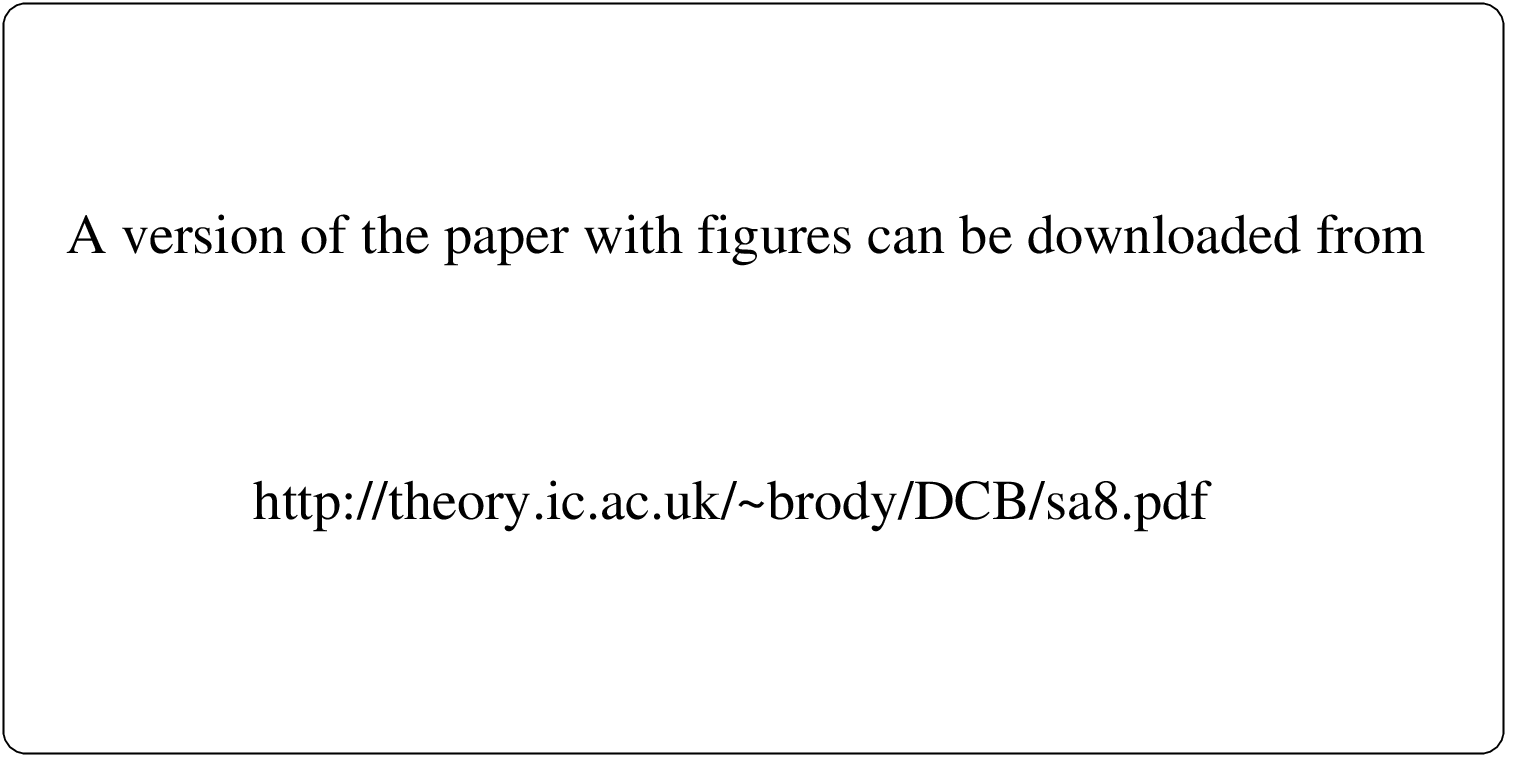,width=12cm,angle=270}}}
\vspace{0cm} \caption{The average of $\rho_t(x)$ for a thousand
runs. The initial and final conditions are given by
$\psi_0(x)=\phi_1(x)$ and $\psi_\infty(x)= \chi_2(x)$,
respectively, and we set $\alpha=2.5$. The $x$-axis is measured in
units of $L$, the time axis is measured in units of
$4\mu^2L^4/\hbar^4\sigma^2$, and the magnitude of $\rho_t(x)$ is
scaled by a factor of 1,000. This plot illustrate how the density
matrix is diagonalised.}
\end{figure}

In order to study the scale of $\tau_R$, we consider the
probability that the process $M_{mj}$ is smaller than ${\rm
e}^{-\lambda}$ for some number $\lambda$. If $\omega_{mj}>0$ this
is given by
\begin{eqnarray}
{\mathbb P}\left(M_{mj}<{\rm e}^{-\lambda}\right) &=& {\mathbb
P}\left( \sigma \omega_{mj}B_t - \half\sigma^2\omega_{mj}^2 t <
-\lambda \right) \nonumber \\ &=& {\mathbb P}\left( B_t < \half
\sigma\omega_{mj}t - \frac{\lambda}{\sigma\omega_{mj}} \right)
\nonumber
\\ &=& N\left( \half \sigma|\omega_{mj}|\sqrt{t} -
\frac{\lambda}{\sigma|\omega_{mj}| \sqrt{t}} \right) .
\label{eq:22}
\end{eqnarray}
Here $N(x)$ is the standard normal distribution function
\begin{eqnarray}
N(x) = \frac{1}{\sqrt{2\pi}} \int_{-\infty}^{x} {\rm
e}^{-\frac{1}{2}y^2} {\rm d}y . \label{eq:23}
\end{eqnarray}
Note that in obtaining the second expression on the right-hand
side of (\ref{eq:22}) we have used the fact that $\omega_{mj}>0$;
whereas if $\omega_{mj}<0$ we would instead have the probability
\begin{eqnarray}
{\mathbb P}\left(M_{mj}<{\rm e}^{-\lambda}\right) =  {\mathbb
P}\left( B_t > \half \sigma\omega_{mj}t - \frac{\lambda}
{\sigma\omega_{mj}} \right) . \label{eq:22-1}
\end{eqnarray}
However, owing to the symmetry relation $N(-x)=1-N(x)$ satisfied
by the normal distribution function, the final result
(\ref{eq:22}) is unaltered.

Now, for relaxation, we would like the probability (\ref{eq:22})
to be sufficiently great, say,
\begin{eqnarray}
{\mathbb P}\left(M_{mj}<{\rm e}^{-\lambda}\right)\geq 0.95,
\label{eq:0.95}
\end{eqnarray}
for a suitably large $\lambda$. The choice of $\lambda$
corresponds to how far the reduction has to proceed for relaxation
to have effectively set in. Because $N(x)\sim 0.95$ when
$x\sim1.65$, we can put a bound on the time variable such that the
probability (\ref{eq:22}) is greater than $0.95$. This is given by
\begin{eqnarray}
\half \sigma|\omega_{mj}|\sqrt{t} - \frac{\lambda}
{\sigma|\omega_{mj}| \sqrt{t}} > 1.65, \label{eq:24-0}
\end{eqnarray}
or equivalently
\begin{eqnarray}
\sqrt{t} > \frac{3.3+\sqrt{3.3^2+8\lambda}}{2\sigma |\omega_{mj}|}
. \label{eq:24}
\end{eqnarray}
In particular, for a fixed value of $j$, we would like this
relation to hold for all values of $m\neq j$. That is to say, we
would like $M_{mj}$ to become negligible for all $m\neq j$, which
will ensure that $H_t^j\rightarrow E_j$. This can be guaranteed by
choosing $m=j+1$, since this choice maximises the right hand side
of (\ref{eq:24}). For example, if we take $\lambda=10$, then the
relaxation timescale obtained is of the order
\begin{eqnarray}
\tau_R \sim \frac{40\alpha^4}{\pi^4 \sigma^2 (2j+1)^2}
\label{eq:25}
\end{eqnarray}
when the terminal state is $\chi_j(x)$. Thus provided
$t\geq\tau_R$, we can be $95\%$ confident that $M_{mj}<10^{-5}$
for every $m\neq j$.

We note, incidentally, that the result in (\ref{eq:25}) is based
on the assumption that the expansion factor $\alpha$ is strictly
and sufficiently greater than one. In the exceptional case where
$\alpha=1$, we have $\tau_R=0$, which follows directly from
(\ref{eq:21}) if one notes that $\pi_m=0$ for every $m\neq j$, and
thus $H_t^j=E_j$ irrespective of the values of $M_{mj}$ and $t$.

When $\alpha\sim1$, the foregoing analysis on the relaxation
timescale needs to be carried out more carefully. For this
purpose, let us consider the case in which we have
$\alpha=1+\epsilon$, where the value of $\epsilon\geq0$ can be
very small. Now, if the system is initially in the ground state
$\phi_1(x)$, then for $\epsilon\ll1$, the transition probability
$\pi_m(\epsilon)$ admits the expansion
\begin{eqnarray}
\pi_m(\epsilon) &\sim& \frac{4\sin^2(\pi m)}{\pi^2(m^2-1)^2} +
\left( \frac{16\sin^2(\pi m)}{\pi^2(m^2-1)^3} \right. \nonumber
\\ & &\ + \left. \frac{4(3 \sin^2(\pi m)-2\pi m\sin(\pi m)\cos(\pi m))}
{\pi^2(m^2-1)^2}\right) \epsilon + \cdots \label{eq:26}
\end{eqnarray}
for all $m$. In particular, $\pi_1(\epsilon)$ is of order one,
whereas the next largest transition probability when $\alpha\sim1$
is given, up to order $\epsilon^2$, by $\pi_2(\epsilon)=
\frac{16}{9}\epsilon^2$. The implication of this is that the
condition for $\pi_m M_{m1}<{\rm e}^{-\lambda}$ to be valid for
all $m\neq1$ with $95\%$ confidence-level is automatically
satisfied for all $t\geq0$, and thus it is not a sufficient
criteria to determine the relaxation timescale.

Therefore, it is important in the small perturbation regime to
determine how small $\epsilon$ can be for the $(p\times100)\%$
confidence-level analysis to be viable. This can be determined if
we replace $\lambda$ in (\ref{eq:22}) by
$\lambda+\ln\pi_m(\epsilon)$ and $0.95$ in (\ref{eq:0.95}) by $p$.
The latter is equivalent to replacing $1.65$ in (\ref{eq:24-0}) by
$N^{-1}(p)$, where $N^{-1}(x)$ is the inverse of the normal
distribution function (\ref{eq:23}). The result gives us
\begin{eqnarray}
\sqrt{t} > \frac{2N^{-1}(p)+ \sqrt{(2N^{-1}(p))^2+8(\lambda+
\ln\pi_m(\epsilon))}}{2\sigma |\omega_{mj}|} , \label{eq:27}
\end{eqnarray}
and we maximise the right-hand side of (\ref{eq:27}) over all
$m\neq j$, when $j=1$. The maximum is obtained by setting $m=2$,
which leads us to the timescale for relaxation into ground state.
This is given by
\begin{eqnarray}
\tau_R \sim \frac{1}{\sigma^2\omega_{21}^2} \left( N^{-1}(p) +
\sqrt{2\lambda + 4\ln(4\epsilon/3)+\left(N^{-1}(p)\right)^2}
\right)^2 . \label{eq:28}
\end{eqnarray}
In particular, for a fixed $\epsilon$, either $\lambda$ or $p$
must be sufficient large so that the right-hand side of
(\ref{eq:28}) is real. Conversely, for fixed $\lambda$ and $p$ the
confidence-level analysis is viable only when $\epsilon$ is large
enough to ensure the reality of the right-hand side of
(\ref{eq:28}).

\section{Towards the quantum adiabatic theorem}

The analysis of the relaxation timescale shows that for an
infinitesimal expansion of potential well the initial eigenstate
$\phi_n(x)$ will almost immediately relax into the corresponding
eigenstate $\chi_n(x)$. This result suggests that when the
potential well is expanded sufficiently slowly, then the dynamics
of relaxation will force the system to {\sl remain} in the $n$-th
eigenstate. In other words, the stochastic evolutionary law
appears to give rise to an analogue of quantum adiabatic theorem.
In what follows we shall explore this idea in greater depth.

For this purpose we need to consider now the dynamics of a quantum
system in the presence of a time-dependent Hamiltonian. Let us
denote by ${\hat H}(t)$ a generic time-dependent Hamiltonian, and
for each fixed time $t$ we write $\chi_n(t,x)$ $(n=1,2,\ldots)$
for the $n$-th eigenstate of the operator ${\hat H}(t)$ with
eigenvalue $E_n(t)$. Now, for an adiabatic approximation, we shall
assume that ${\hat H}(t)$, $\chi_n(t,x)$, and  $E_n(t)$ are
continuous in $t$, and vary sufficiently slowly so that we can
write
\begin{eqnarray}
\frac{\partial\chi_n(t,x)}{\partial t} \approx -{\rm i} E_n(t)
\chi_n(t,x). \label{eq:8.1}
\end{eqnarray}
Then for a general state $\psi_t(x)$ the Schr\"odinger evolution
can be written as
\begin{eqnarray}
\frac{\partial\psi_t(x)}{\partial t} = -{\rm i} {\hat H} (t)
\psi_t(x) = -{\rm i} {\hat H}(t) \sum_{n} a_n(t) \chi_n(t,x) .
\label{eq:8.2}
\end{eqnarray}
Therefore, we find that
\begin{eqnarray}
\sum_{n} \dot{a}_n(t) \chi_n(t,x) =0 \label{eq:8.3}
\end{eqnarray}
in the adiabatic approximation, from which it follows at once
${\dot a}_k(t)=0$. Thus the solution to the deterministic
Schr\"odinger equation can be written as
\begin{eqnarray}
\psi_t(x) = \sum_{n} a_n \chi_n(t,x) . \label{eq:8.4}
\end{eqnarray}

These relations are the consequences of the adiabatic
approximation (\ref{eq:8.1}) in the deterministic unitary theory.
What we would like to consider here is the implication of the
assumption (\ref{eq:8.1}) in the case of the stochastic
evolutionary law (\ref{eq:3.1}), which in the present context is
given by
\begin{eqnarray}
{\rm d}\psi_t(x) &=& -{\rm i}{\hat H}(t)\psi_t(x){\rm d}t - \octa
\sigma^2 ({\hat H}(t)-H_t)^2 \psi_t(x) {\rm d}t \nonumber \\ & & +
\half \sigma ({\hat H}(t)-H_t)\psi_t(x) {\rm d}W_t, \label{eq:8.5}
\end{eqnarray}
where
\begin{eqnarray}
H_t=\frac{\int \psi_t^*(x){\hat H}(t)\psi_t(x){\rm d}x}{\int
\psi_t^*(x)\psi_t(x){\rm d}x}.
\end{eqnarray}
We note that (\ref{eq:8.5}) preserves the norm of $\psi_t(x)$.
However, for the energy process $H_t$ we have
\begin{eqnarray}
{\rm d}H_t = {\dot H}_t{\rm d}t + \sigma V_t {\rm d}W_t ,
\label{eq:8.6}
\end{eqnarray}
instead of (\ref{eq:2.2}), where
\begin{eqnarray}
{\dot H}_t=\int\psi_t^*(x) (\partial_t{\hat H}(t))\psi_t(x){\rm
d}x.
\end{eqnarray}
It is natural that the process corresponding to the expectation
value of the Hamiltonian is no longer a martingale in the
time-dependent case, but rather exhibits a drift, the sign of
which depends on the expectation value of the time derivative of
the Hamiltonian. Similarly for the variance process we obtain
\begin{eqnarray}
{\rm d}V_t &=& -\sigma^2 V_t^2 {\rm d}t + \sigma \beta_t {\rm
d}W_t \nonumber \\ & & +2 \left( \int\psi_t^*(x)({\hat H}(t)-H_t)
(\partial_t{\hat H}(t)-{\dot H}_t)\psi_t(x){\rm d}x \right) {\rm
d}t. \label{eq:8.7}
\end{eqnarray}

Now, for state reduction, a necessarily condition is that the
variance process $V_t$ is a supermartingale, i.e. on average a
decreasing process. We observe that the last term in
(\ref{eq:8.7}) is given by the covariance of the two operators
${\hat H}(t)$ and $\partial_t{\hat H}(t)$, which can be positive
or negative. If it is negative, then the variance process $V_t$ is
a supermartingale, whereas if it is positive, $V_t$ can still be a
supermartingale, provided the covariance is not too large. In
particular, if we write
\begin{eqnarray}
\Delta H_t=\sqrt{V_t}
\end{eqnarray}
 and
\begin{eqnarray}
\Delta {\dot H}_t=\sqrt{{\rm Var}[\partial_t{\hat H}(t)]}
\end{eqnarray}
respectively for the standard deviations of ${\hat H}(t)$ and
$\partial_t{\hat H}(t)$, then for the supermartingale condition we
require
\begin{eqnarray}
2\rho_t \Delta {\dot H}_t \Delta H_t < \sigma^2 (\Delta H_t)^4 ,
\label{eq:8.8}
\end{eqnarray}
where $\rho_t$ is the correlation between ${\hat H}(t)$ and
$\partial_t{\hat H}(t)$. If we notice the fact that $-1\leq\rho_t
\leq1$, we find, to ensure that $V_t$ is a supermartingale, it
suffices that the following relation should hold:
\begin{eqnarray}
\frac{\Delta{\dot H}_t}{\Delta H_t} < \half\sigma^2 (\Delta H_t)^2
. \label{eq:8.9}
\end{eqnarray}
In other words, we require the uncertainty in the change of ${\hat
H}$ to be small compared with the uncertainty in ${\hat H}$
itself. Intuitively, if the change of the Hamiltonian is
sufficiently slow, then $\partial_t{\hat H}(t)$ will be small, and
therefore we would expect $\Delta{\dot H}_t\ll\Delta H_t$ be valid
in the adiabatic regime. In particular, relation (\ref{eq:8.9})
can be viewed as the consequence of adiabatic motion, which
ensures that the variance $V_t$ is a supermartingale.

Now we turn to the the consideration of the stochastic equation
(\ref{eq:8.5}). To begin with, let us define the process $\Pi_t^k$
by
\begin{eqnarray}
\Pi_t^k = \Pi_0^k \exp\left(\sigma\int_0^t(E_k(s)-H_s){\rm d}W_s -
\half\sigma^2\int_0^t(E_k(s)-H_s)^2{\rm d}s\right) .
\label{eq:8.10}
\end{eqnarray}
Then it is a straightforward exercise in Ito calculus to deduce
that
\begin{eqnarray}
{\rm d}\Pi_t^k=\sigma(E_k(t)-H_t)\Pi_t^k{\rm d} W_t,
\label{eq:8.11}
\end{eqnarray}
from which it follows that the dynamics of the process
\begin{eqnarray}
a_{kt}=(\Pi_t^k)^{1/2}
\end{eqnarray}
is given by
\begin{eqnarray}
{\rm d}a_{kt} = -\octa\sigma^2 (E_k(t)-H_t)^2 a_{kt} {\rm d}t +
\half (E_k(t)-H_t) a_{kt} {\rm d}W_t , \label{eq:8.12}
\end{eqnarray}
Therefore, if we define
\begin{eqnarray}
\psi_t(x) = \sum_j a_{kt} \chi_j(t,x) , \label{eq:8.13}
\end{eqnarray}
where the functions $\chi_j(t,x)$ are eigenfunctions of ${\hat
H}(t)$, then we obtain
\begin{eqnarray}
{\rm d}\psi_t(x) &=& \sum_k \left({\rm d} a_{kt}\right)\chi_k(t,x)
+ \sum_k a_{kt} \left(\partial_t\chi_k(t,x)\right),
\end{eqnarray}
and thus
\begin{eqnarray}
{\rm d}\psi_t(x) &=& -{\rm i}{\hat H}(t)\psi_t(x){\rm d}t - \octa
\sigma^2 ({\hat H}(t)-H_t)^2 \psi_t(x) {\rm d}t \nonumber \\ & & +
\half \sigma ({\hat H}(t)-H_t)\psi_t(x) {\rm d}W_t \nonumber \\ &
& + \left( \sum_k a_{kt}\left(\partial_t\chi_k(t,x)-{\hat H}(t)
\chi_k(t,x) \right) \right){\rm d}t. \label{eq:8.14}
\end{eqnarray}
Thus far our analysis is exact. Now, in the adiabatic regime, we
have the relation (\ref{eq:8.1}), which implies that
\begin{eqnarray}
(\partial_t\chi_k(t,x) \approx{\hat H}(t)\chi_k(t,x).
\end{eqnarray}
Therefore, substituting this relation into equation
(\ref{eq:8.14}), we see that {\it in the adiabatic approximation,
the process $\psi_t(x)$ defined by {\rm (\ref{eq:8.13})} is the
solution to the stochastic dynamics {\rm (\ref{eq:8.5})}}.

We observe that in the adiabatic approximation the process
$\Pi_t^k$ has the interpretation that it represents the
probability that $\psi_t(x)$ is in the eigenstate $\chi_k(t,x)$.
On the other hand, $\Pi_t^k$ is also a martingale, satisfying
(\ref{eq:8.11}). Thus the {\sl expected} probability that
$\psi_t(x)$ is in the state $\chi_k(t,x)$, given information up to
time $s$, is precisely the probability that $\psi_s(x)$ is in the
state $\chi_k(s,x)$. It follows, therefore, that if $\psi_t(x)$ is
in the state $\chi_k(0,x)$ at time $0$, it will be in the state
$\chi_k(t,x)$ at time $t$ with probability one.

\section{Discussion}

Although in the present paper we have primarily examined the
square-well potential, many of the results obtained, namely, the
wave function (\ref{eq:16}), the conditional Hamiltonian process
(\ref{eq:18}), the probability distribution of the particle
(\ref{eq:19}), and the bound on relaxation time (\ref{eq:24}), are
independent of the specific model being considered. As long as the
eigenvalues $E_m$ and the eigenfunctions $\chi_m(x)$ of the
Hamiltonian are known, either analytically or numerically, these
results can be applied to study the details of the dynamics. In
other words, the general approach outlined here, based on the
nonlinear filtering methodology introduced in Brody $\&$ Hughston
(2002), can be applied to study a wide range of problems in
perturbation theory, including of course a more generic
time-dependent Hamiltonian. We hope to take up this line of
investigation in greater detail elsewhere.

The point of view we have put forward in this paper is that the
dynamical law (\ref{eq:3.1}) offers a simple but plausible
characterisation of the subsequent evolution of a quantum system
after the Hamiltonian has been perturbed. If a measurement of the
system energy is carried out after a passage of time greater than
the relaxation timescale, then according to the stochastic
postulate, when the system is initially in the eigenstate
$\phi_n(x)$, the eigenvalue $E_m$ will be observed with
probability $\pi_{nm}$. This is because the system is already in
the eigenstate $\chi_m(x)$ with that probability. On the other
hand, according to the unitarity postulate associated with the
Schr\"odinger equation, the system is in a superposition of a
myriad of eigenstates. Nevertheless, the outcome of the energy
measurement will give $E_m$ with probability $\pi_{nm}$. The
question thus arising is whether one can distinguish the two
theories by means of a suitable experiment. To this end, we note
that, for time $t\geq \tau_R$, the state of the system, according
to the stochastic law, is given by a mixed-state density matrix,
whose diagonal elements, in the energy basis, are given by
$\pi_m=\pi_{nm}$; whereas according to unitary law, the system is
in a pure state, given by the superposition of the energy
eigenstates, whose coefficients are given by $\sqrt{\pi_m}$.
Therefore, if these pure and mixed states can be distinguished
statistically in some way by means of experiments, then in
principle one can rule out at least one of the two postulates
indicated here. At present, the possibility of distinguishing the
two remains an open problem.

\vspace{0.5cm}
\begin{footnotesize}
\noindent The authors are grateful to S.~L.~Adler, J.~Anandan,
A.~Bassi, T.~A.~Brun, I.~Buckley, and B.~K.~Meister for
stimulating discussions. DCB acknowledges support from The Royal
Society. LPH acknowledges the support and hospitality of the
Institute for Advanced Study, where part of this work was carried
out.
\end{footnotesize}

\end{document}